\documentclass[a4paper,12pt]{article}
\usepackage[english]{babel}
\usepackage{latexsym}
\date{}
\begin{document}
\title
{GENERAL RELATIVITY, \\ GRAVITATIONAL ENERGY AND SPIN--TWO FIELD}
\author
{Leszek M. SOKO\L{}OWSKI\thanks{E-mail: UFLSOKOL@TH.IF.UJ.EDU.PL} \\
Astronomical Observatory and Centre for Astrophysics,\\
Jagellonian University, \\
Orla 171, Krak\'ow 30-244, Poland}
\maketitle
\begin{abstract}
In my lectures I will deal with three seemingly unrelated problems: 
i) to what extent is general relativity exceptional among metric 
gravity theories? ii) is it possible to define gravitational energy 
density applying field--theory approach to gravity? and iii) can a 
consistent theory of a gravitationally interacting spin--two field be 
developed at all? The connecting link to them is the concept of a 
fundamental classical spin--2 field. A linear spin--2 field introduced as a 
small perturbation of a Ricci--flat spacetime metric, is gauge 
invariant while its energy--momentum is gauge dependent. Furthermore, 
when coupled to gravity, the field reveals insurmountable inconsistencies 
in the resulting equations of motion. 
After discussing the inconsistencies of any coupling of the linear spin--2 
field to gravity, I exhibit the origin of the fact that a gauge invariant 
field has the variational metric stress tensor which is gauge dependent. I 
give a general theorem explaining under what conditions a symmetry of a 
field Lagrangian becomes also the symmetry of the variational stress tensor. 
It is a conclusion of the theorem that any attempt to define gravitational 
energy density in the framework of a field theory of gravity must fail. 
Finally I make a very brief introduction to basic concepts of how a certain 
kind of a necessarily nonlinear spin--2 field arises in a natural way from 
vacuum nonlinear metric gravity theories (Lagrangian being any scalar 
function of Ricci tensor). This specific spin--2 field consistently 
interacts gravitationally and the theory of the field is promising. 
\end{abstract}


\subsection*{1. Introduction}
General relativity is just a point in the "space" of all existing 
and conceivable theories of gravitational interactions. Nevertheless 
all the theories other than Einstein's one, named "alternative theories 
of gravity", have rather bad reputation among most relativists: general 
relativity is enough complicated in itself and well confirmed by all known 
empirical data so that there seems no point in considering more intricate 
theories whose confirmation is either rather poor or presently non--existing 
at all. In fact, the alternative theories are some generalizations of 
general relativity, which is invariably used as a reference point for their 
construction. These modifications go in all possible directions making any 
attempt to confront them with experiment very difficult. \\
  In spite of this criticism in the last three decades there has been 
considerable interest in some alternative theories. It comes from various 
directions but the generic attitude in it is to search for a deeper relationship 
between gravitational physics and elementary particle interactions. Quantum 
gravity and superstrings seem to indicate that definite modifications of 
general relativity are necessary. Though these fundamental theories are still 
rather fancy than a fact, they suggest that general relativity should be 
replaced in high energy approximation by an effective theory of gravity, 
which should be some metric nonlinear gravity (NLG) theory and it is here 
that a significant progress has been made since early 1980's. (Similar 
methods of investigation can be applied to purely affine and metric--affine 
gravity theories which are presented in other lectures). In NLG theories 
Einstein--Hilbert Lagrangian (the curvature scalar $R$) is replaced by any 
smooth scalar function of Riemann tensor, $L=f(R_{\alpha\beta\mu\nu}, 
g_{\mu\nu})$. All other axioms of general relativity hold for these theories. 
(General relativity and NLG theories may be studied in any spacetime dimension 
$d\geq 4$. In these lectures I will assume $d=4$ unless otherwise is stated.) 
For $f$ is arbitrary, the NLG theories form a densely populated neighbourhood 
of general relativity in the space of gravity theories. This raises a 
fundamental problem: is Einstein theory merely a point of this neighbourhood? 
In other terms: is general relativity distinguished merely by tradition and 
(relative) computational simplicity or does it occupy a preferred position 
with respect to the theories that surround it? It was shown by Magnano, Ferraris 
and Francaviglia \cite{MFF1} and independently by Jakubiec and Kijowski 
\cite{JK} that \emph{NLG 
theories are mathematically, i.e.\/ dynamically equivalent to general 
relativity and there are convincing arguments that at least some of them are 
also physically equivalent to it}. In a sense these theories represent 
Einstein theory in disguise. General relativity is not surrounded by 
theories different from it and its closest neighbourhood consists of its own 
versions in distinct (usually unphysical) variables. General relativity is 
an isolated point in the space of gravity theories. Physical interpretation 
of these versions of Einstein theory is however quite subtle, see \cite{MS1}.
\\

If general relativity preserves its distinguished and leading role, one should 
return to unsolved problems of this theory, e.g.\/ to that of localizability 
of gravitational energy. A conventional wisdom claims that gravitational 
energy and momentum densities are nonmeasurable quantities since the 
gravitational field can always be locally transformed away according to 
Strong Equivalence Principle. Nevertheless since the very advent of general 
relativity there have been numerous attempts to construct a local concept of 
gravitational energy as it would be very useful in dealing with practical 
problems, e.g.\/ a detailed description of cosmological preturbations in the 
early universe. Among various approaches to the problem a particularly 
promising one has been provided by the field theory formulation of gravity 
theory, according to which gravity is just a tensor field existing in 
Minkowski space and the latter may or may not be regarded as the spacetime of 
the physical world. In this formulation gravity is described by a Lagrangian 
field theory in flat spacetime for a spin--two field and its energy density 
is given by the metric (i.e.\/ variational) energy--momentum tensor. In 
linearized around Minkowski space general relavity a linear spin--2 field 
arises and 
it is gauge invariant while its metric energy momentum tensor is not. Thus 
a naive field-theoretical approach to gravity fails. It is interesting on its own 
to explain why the energy--momentum tensor for the field does not inherit 
the symmetry property of the underlying Lagrangian. Another problem, closely 
related to the previous one, is whether can this defect be overcome at all: 
is there a linear spin--2 field which is dynamically equivalent to linearized 
gravity and possessing a gauge invariant energy--momentum tensor? The answer 
is no and no such field exists because the Lagrangian of the linearized general 
relativity is gauge invariant only in empty spacetimes ($R_{\mu\nu}=0$). In 
general, the energy-momentum tensor does not inherit a symmetry of the 
Lagrangian if the symmetry does not hold in a generic curved spacetime. \\

This failure does not cancel interest in spin--2 fields since a certain kind of 
necessarily nonlinear spin--2 field arises in a natural way from vacuum NLG 
theories as a component of a multiplet of fields describing gravity. Any 
Lagrangian different from $R$ and the Euler-Poincar\'e topological invariant 
density (Gauss--Bonnet term) gives rise to fourth order field equations and 
hence gravity has altogether eight (rather than two) degrees of freedom. The 
"particle content" of these degrees of freedom is disclosed by decomposing 
the gravitational field into a multiplet of fields with definite spins. The 
decomposition is accomplished by using a specific Legendre transformation. 
One gets a triplet: spacetime metric (2 degrees of freedom), a scalar field 
(1 d.o.f.) and massve spin--2 field (5 d.o.f.). The Legendre transformation 
reduces 
the fourth--order theory to a second--order one. Actually there are two 
different Legendre transformations giving rise to two different (dynamically 
equivalent) versions of the resulting theory. Both versions take on the form 
of Einstein's theory for the metric field with the other two fields of the 
triplet acting as a "matter" source in Einstein field equations. For a 
special form of the original Lagrangian for NLG theory the scalar field 
disappears and the propagation of the metric field is governed by the nonlinear 
spin--2 field alone. Thus any NLG theory generates a consistent gravitational 
interaction for the nonlinear massive spin--2 field (in general also coupled 
to the scalar). \\

The latter outcome is quite unexpected and surprising since it has been known 
for more than three decades that a linear (and gauge invariant) spin--2 field 
cannot interact gravitationally in a consistent way, it can only propagate as 
a test field in an empty spacetime. For higher spin fields the situation is 
even worse: it was found that in order for these fields 
to exist on a Lorentzian manifold, restrictions on the curvature, being in 
fact consistency conditions, were required. These conditions are very 
severe: "there is no easy way to have physical fields with spins $> 2$ on 
anything but a flat manifold" \cite{ChD}. It is therefore a common belief 
among field theorists that
Nature avoids the consistency problem by simply not creating fundamental
spin--two (nor higher spin) fields except gravity itself. This is, however, 
not true and consistent dynamics for spin--2 fields in the framework of general 
relativity is provided by NLG theories; the fields must be nonlinear and their 
theory is built up in a way quite different from that for the linear (and 
inconsistent) field. \\

In these lectures I will follow the opposite order to that presented above. I 
will start from linear spin--2 field in flat spacetime and show that it is 
unphysical in the sense that it cannot exist (i.e. consistently interact 
gravitationally) in non--empty spacetimes. Then I will prove that no linear 
spin--2 field can have a gauge invariant energy--momentum tensor in flat 
spacetime. Finally I will discuss NLG theories and show how they generate 
the nonlinear massive spin--two field. 

\subsection*{2. Linear free spin-2 field in Minkowski space}
It is commonly accepted that a matter field of integer spin $s$ is 
described by a symmetric tensor field with $s$ indices, $\psi_{\mu_1
\ldots\mu_s}$. Thus for $s=2$ it is described by a tensor 
$\psi_{\mu\nu}=\psi_{\nu\mu}$. We 
shall focus our attention on spin-2 fields in this tensor 
representation\footnote{In a 4-dimensional spacetime there exists 
another representation 
of a \emph{linear\/} spin two: it is mathematically described by a 
4-index tensor field $H_{\alpha\beta\mu\nu}$ having all algebraic 
symmetries of Riemann tensor \cite{AD1}. Then the tensor 
$\psi_{\mu\nu}$ is just a superpotential for the field 
$H_{\alpha\beta\mu\nu}$. Since this alternative mathematical description 
cannot be 
used for the nonlinear spin-2 fields arising in NLG theories, we 
infer that the description in terms of symmetric tensors $\psi_{\mu
\nu}$ is more fundamental and we shall not deal with the "Riemann 
tensor-like" representation of spin two.}. Whether the tensor $\psi$ 
represents a gauge-dependent tensor potential or a measurable field 
strength depends on its dynamical properties. Assuming 
four-dimensionality of the spacetime we define spin (or helicity for 
massless matter fields) by the transformation properties of a tensor 
field under the rotation group in a three-dimensional space. Then 
a spin-2 field should have $2s+1=5$ degrees of freedom. On the other 
hand the symmetric tensor $\psi_{\mu\nu}$ has 10 algebraically 
independent components. Thus \emph{any\/} equations of motion should 
be compatible with a system of algebraic (and differential) constraints 
removing 5 spurious unphysical components ("unphysical modes"). In the 
case of massless fields of helicity two there are only two degrees of 
freedom and constraint equations should set to zero eight unphysical 
modes. In general, 
for spin $s\geq 1$ equations of motion must necessarily be supplemented 
by a number of constraints. This fact makes the dynamics of $s\geq 1$ 
fields rather complicated and for $s\geq 2$ excludes a consistent 
interaction of a linear field with gravity. \\

There are four stages of constructing a theory for interacting fields. \\
i) Choice of a mathematical object (field) describing classical matter 
carrying spin $s\geq 1$. As mentioned above we shall describe any 
continuous spin-2 matter by a symmetric tensor field $\psi_{\mu\nu}$. \\
ii) Construction of free field theory in flat spacetime. The theory is 
first built for a massless field and the guiding principle is the 
postulate of a gauge invariance. Then for a massive field an 
appropriate mass term is added to the Lagrangian. The free field theory 
is consistent (after taking into account all the existing constraints). \\
iii) Construction of self-interaction for the field. It is here that 
some troubles may arise. We shall not deal with this case. \\
iv) Coupling of the field to other kinds of matter or to gravity. 
A genuine inconsistency arises when one attempts 
to introduce gravitational interactions for spin two. \\

To obtain linear equations of motion one assumes a Lagrangian quadratic 
in the field potential or field strength. Taking the variational derivative 
with respect to the 10 algebraically independent components of $\psi_{\mu
\nu}$ one arrives at a system of 10 Lagrange equations, while, there are 
at most 5 physical degrees of freedom (in 
the massive case) and some components are redundant (represent a "pure 
gauge"). Then Lagrange field equations form a \emph{degenerate system\/}:
not all of them are second order hyperbolic (i.e. propagation) equations 
and a number of them are first order constraints on the initial 
data---they do not determine ${\partial^2 \over \partial t^2}
\psi_{\mu\nu}$ for some components of $\psi$ in terms of arbitrary initial 
data and actually represent restrictions on the data. These constraints 
will be referred to as \emph{primary constraints\/}. Then applying various 
linear operations to Lagrange equations one can generate from them a 
number of \emph{secondary constraints}\footnote{We do not define here 
the difference between primary and secondary constraints in a precise 
way. By primary constraints we mean the ones recognized in the system of 
field equations just by inspection (or possibly by taking a linear 
combination of the equations); secondary constraints are those generated 
by applying various differential operators to the system.}. The secondary 
constraints allow one to transform the original system of Lagrange 
equations into a nondegenerate system of hyperbolic propagation equations 
for which the Cauchy problem is well posed.\\
It turns out that finding out the appropriate Lagrangian is not 
straightforward and Fierz and Pauli \cite{FP} who first did it in 1939 had 
to use a rather indirect procedure including introducing at an 
intermediate stage an auxiliary unphysical field which was set to 
zero at the end. The Lagrangian is quite complicated in comparison 
to that for a vector field and may be given in a number of equivalent 
(up to a total divergence term) forms. \\
As we are interested in gravitational interactions of  
spin-2 field, we omit the free field theory in flat spacetime and construct a 
Lagrangian for $\psi_{\mu\nu}$ in a curved spacetime.

\subsection*{3. Linear spin-2 field in a curved spacetime}
  The best way of constructing a theory for gravitationally interacting 
spin--2 field, massive or massless, is just to employ a gravitational 
perturbation analogy \cite{AD2}. One takes any 
spacetime metric and perturbs it, 
$g_{\mu\nu}\rightarrow g_{\mu\nu}+\delta g_{\mu\nu}$. 
The second variation of Einstein-Hilbert 
action $S[g]=\int d^4x\sqrt{-g}R$ evaluated at the 
"background" metric $g_{\mu\nu}$ is a functional quadratic in 
the metric fluctuations and if one identifies $\delta g_{\mu\nu}$ 
with $\psi_{\mu\nu}$ the functional provides an action giving 
rise to linear Lagrange equations for the field. Actually 
there is a considerable freedom in constructing a Lagrangian for the field 
giving rise to various inequivalent models. Different models arise if the 
variable to be varied is not the the metric but a function of it. 
As a result the functional dependence of $\delta^2 S$ on the variations of 
the chosen variable is different in each case, while it is always chosen 
that $\psi_{\mu\nu}$ is equal to the metric perturbation. The Lagrangians 
and corresponding field equations differ by a number of terms involving 
Ricci tensor but not Weyl tensor. As a consequence all these models become 
equivalent in empty spacetime, $R_{\mu\nu}=0$. Most of the models are 
nonminimal in the sense that they assume direct coupling of the field to 
Ricci tensor in their Lagrangians. And all models suffer the same 
deficiences whenever $R_{\mu\nu}\ne 0$, i.e.\/ when there is a source of 
gravity and these defects preclude a consistent theory of a linear 
purely spin--2 field in a generic spacetime. Here we give examples of two 
models. \\

The simplest choice of the variable to be varied is the metric $g^{\mu\nu}$ 
itself, then $\psi_{\mu\nu} =\delta g_{\mu\nu} = - g_{\mu\alpha}
g_{\nu\beta} \delta g^{\alpha\beta}$. The action for $\psi_{\mu\nu}$  
is defined as $S[\psi]\equiv \frac{1}{2}\delta(\delta S[g])$.
Next one assumes that 
$\psi_{\mu\nu}$ is a non-geometric tensor field which 
interacts with gravity. In other words from now on $g_{\mu\nu}$ is 
not regarded as a fixed background metric but rather as a dynamical 
field coupled to $\psi_{\mu\nu}$. (This means that the equations of 
motion for the fields are {\em not\/} perturbation equations of a 
given solution for pure gravity.) To assign a mass to $\psi_{\mu\nu}$ 
one puts in the Lagrangian by hand a mass term which is appropriately 
chosen to avoid any additional scalar field. After making these 
operations a second 
order action for $\psi_{\mu\nu}$ reads (the action is linear in 
second derivatives) 
$$S[\psi]=-\frac{1}{2}\int d^4x\sqrt{-g}[\psi^{\mu\nu}G^L_{\mu\nu}
+{1\over 2}\psi\psi^{\mu\nu}G_{\mu\nu}+{m^2\over 
2}(\psi_{\mu\nu}\psi^{\mu\nu}-\psi^2)]\eqno(1)$$ 
where $\psi = g^{\mu\nu}\psi_{\mu\nu}$ and $G^L_{\mu\nu}$ is 
the linear in $\psi_{\mu\nu}$ part of Einstein tensor, 
$$G_{\mu\nu}(g+\psi)=G_{\mu\nu}(g)+G^L_{\mu\nu}(\psi, g)+\ldots,$$
and 
\setcounter{equation}{1}  
\begin{eqnarray}
&
G^L_{\mu\nu}(\psi, g)\equiv {1\over 2}(-\Box\psi_{\mu\nu}+ 
\psi^{\alpha}_{\;\;\mu ;\nu\alpha}+\psi^{\alpha}
_{\;\;\nu ;\mu\alpha}-\psi_{;\mu\nu}-g_{\mu\nu}\psi^{\alpha\beta}_
{\;\;\; ;\alpha\beta}+g_{\mu\nu}\Box\psi+
&
\nonumber\\
&
g_{\mu\nu}\psi^{\alpha
\beta}R_{\alpha\beta}-\psi_{\mu\nu}R);
&
\end{eqnarray}
here $\Box T \equiv T_{;\alpha}^{\;\; ;\alpha}$. 
Adding Einstein-Hilbert action $S[g]$ to $S[\psi]$ one derives the 
full system of equations of motion consisting of Einstein's field 
equations $G_{\mu\nu}=T_{\mu\nu}(\psi, g)$ (we set $8\pi G=c=\hbar=
1$) where $T_{\mu\nu}$ is the variational energy-momentum tensor 
following from (1) and Lagrange equations 
$$ G^L_{\mu\nu}(\psi, g)+{1\over 2}G_{\mu\nu}(g)
\psi-{m^2\over 2}(g_{\mu\nu}\psi-\psi_{\mu\nu})=0.\eqno(3)$$ 
One sees that the simplest choice of the varied variable leads to a nonminimal 
model. We shall not investigate further this model and consider a minimal one.\\

A minimal model, which is unique up to field redefinitions, arises if the 
independent variable to be varied is the tensor density 
$\tilde{g}^{\mu\nu}\equiv \sqrt{-g} g^{\mu\nu}$. The first variation 
of $S$ is the same as in the nonminimal case presented above, 
$\delta_{\tilde{g}} S =\delta S$. A difference appears in  
evaluating the second variation, for, by definition, $\delta_{\tilde{g}}  
\delta \tilde{g}^{\mu\nu}\equiv 0$ while $\delta \delta \tilde{g}^{\mu\nu}
\ne 0$ due to the relationship 
$$\delta \tilde{g}^{\mu\nu} = \sqrt{-g}(-\frac{1}{2} g^{\mu\nu}g_{\alpha\beta}
\delta g^{\alpha\beta} + \delta g^{\mu\nu}).$$
The final outcome is a second order action for the field $\psi_{\mu\nu}$ 
which is related to $\delta \tilde{g}^{\mu\nu}$ according to the preceding 
equation by 
$$\delta \tilde{g}^{\mu\nu} = -\sqrt{-g}(\psi^{\mu\nu}
-\frac{1}{2} g^{\mu\nu}\psi);$$
the action reads
\setcounter{equation}{3}  
\begin{eqnarray}
&
S_{II}[\psi]= -\frac{1}{4}\int d^4x\sqrt{-g}[\psi^{\mu\nu}
(-\Box\psi_{\mu\nu}+ 
\psi^{\alpha}_{\;\;\mu ;\nu\alpha}+\psi^{\alpha}
_{\;\;\nu ;\mu\alpha}-\psi_{;\mu\nu}-
&
\nonumber\\
&
g_{\mu\nu}\psi^{\alpha\beta}_
{\;\;\; ;\alpha\beta}+g_{\mu\nu}\Box\psi)+
m^2 (\psi_{\mu\nu}\psi^{\mu\nu}-\psi^2)]
&
\nonumber\\
&
=-\frac{1}{2}\int d^4x\sqrt{-g}[\psi^{\mu\nu} G^{AD}_{\mu\nu}(\psi,g)
+{m^2\over 2}(\psi_{\mu\nu}\psi^{\mu\nu}-\psi^2)].
\end{eqnarray}
Here $G^{AD}_{\mu\nu}(\psi,g)$ is the curvature free part of 
$G^L_{\mu\nu}(\psi,g)$, i.e. 
$$G^{AD}_{\mu\nu}(\psi,g) \equiv G^L_{\mu\nu}(\psi,g)-\frac{1}{2}
(g_{\mu\nu}R^{\alpha\beta}\psi_{\alpha\beta} - R\psi_{\mu\nu}).
\eqno(5)$$
Hence the action (4) is the curvature free part of the action (1) for the 
nonminimal model. We stress that in both models the tensor field 
$\psi_{\mu\nu}$ (which should be a purely spin--2 field) is formally defined 
as a metric perturbation. One can also replace $S_{II}$ by a first order action, to 
this end one expresses terms with second derivatives by appropriate terms 
made up of first derivatives of the field plus divergences which are 
discarded. 
The resulting action has an unambiguous form in any curved spacetime, 
$$S_W[\psi] = \int d^4x\sqrt{-g}[L_W(\psi,g) - \frac{m^2}{4} 
(\psi_{\mu\nu}\psi^{\mu\nu}-\psi^2)]\eqno(6)$$
with
$$L_W(\psi,g)=\frac{1}{4}\left(-\psi^{\mu\nu;\alpha}\psi_{\mu\nu;\alpha}
+2\psi^{\mu\nu;\alpha}\psi_{\alpha\mu;\nu}
-2\psi^{\mu\nu}{}_{;\nu}\psi_{;\mu}+\psi^{;\mu}\psi_{;\mu}\right).
\eqno(7)$$
This Lagrangian appeared first in the textbook \cite{W} and
will be referred hereafter to as \emph{Wentzel Lagrangian}. Actually in 
Minkowski space the choice of a
Lagrangian for $\psi_{\mu\nu}$ is not unique and a number of equivalent 
forms exist. For example
one can replace the second term in (7) by a more symmetric term
$\psi^{\mu\nu}{}_{;\nu}\psi_{\mu\alpha}{}^{;\alpha}$ and the resulting 
Lagrangian $L_S$ differs from
$L_W$ by a full divergence. However in a curved spacetime the two 
Lagrangians become inequivalent as they differ by a curvature term,
$L_S(\psi,g) = L_W+\mathrm{div}+H$, 
where $H\equiv \psi^{\nu}{}_{\mu}\psi^{\mu\alpha}{}_{;[\nu\alpha]}=
\frac{1}{2}\psi^{\alpha\beta}(\psi_{\beta}{}^{\mu} R_{\mu\alpha} +
\psi^{\mu\nu}R_{\mu\alpha\beta\nu})$. We shall use $L_W$ or its second order 
version (4). \\
For simplicity and possible physical relevance we shall investigate 
only the massless field; in the massive case the final conclusions are 
similar.

\subsection*{4. Massless field, gauge invariance and consistency}
The action functionals (4) and (6) generate the variational 
energy--momentum tensor for the field (we shall deal with it below) 
and Lagrange equations (for $m=0$)
$$E_{\mu\nu} = - G^{AD}_{\mu\nu}(\psi,g) =0.\eqno(8)$$
These form a degenerate system: only 6 out of 10 
equations $E_{\mu\nu}=0$ are hyperbolic propagation ones for 
$\psi_{\mu\nu}$, four equations $E_{0\mu}(g, \psi)=0$ do not 
contain second time derivatives of the field and constitute  
primary constraints on the initial Cauchy data. There are no other 
primary constraints. To get a 
consistent dynamics one should replace the primary constraints by 
secondary constraints which allow to transform the primary ones in 
four missing propagation equations. In flat spacetime the procedure of 
deriving secondary constraints is different in the massive and massless 
case, while it turns out that in presence of curvature the situation is 
somewhat mixed. In Minkowski space the linear tensor $G^L_{\mu\nu}(\psi)$ 
coincides with$G^{AD}_{\mu\nu}(\psi)$ and it is well known from the 
theory of linear perturbations that the former satisfies the linearized 
version of Bianchi identity (in Cartesian coordinates)  
$$\partial^{\nu}G^L_{\mu\nu}(\eta, \psi)\equiv 0.\eqno(9)$$
Yet in a generic spacetime divergence of the field equations does not 
identically vanish, 
$$\nabla^{\nu}E_{\mu\nu}(\psi, g)\equiv Q_{\mu}(\psi, g)=
(-R_{\mu\alpha ;\beta}+{1\over 2}R_{\alpha\beta ;\mu}- R_{\mu\alpha}
\nabla_{\beta}+{1\over 2}g_{\alpha\beta}R_{\mu}^{\;\;\nu}\nabla
_{\nu})\psi^{\alpha\beta}=0,
\eqno(10)$$ 
to derive this expression one has applied the linearized Bianchi identity 
$$\nabla^{\nu}G^L_{\mu\nu}(\psi, g)\equiv \nabla_{\nu}(\psi
^{\nu\alpha}G_{\mu\alpha})+{1\over 2}\psi^{\alpha\beta}_{\;\;\; 
;\mu}G_{\alpha\beta}-{1\over 2}\psi_{;\alpha}G^{\alpha}_{\;\;\mu}.
\eqno(11)$$ 
These are four secondary constraints for $\psi_{\mu\nu}$ in presence 
of gravitation. This outcome resembles the massive case in flat 
spacetime. However in flat spacetime the scalar equation arising as 
the trace of the field equations, $\eta^{\mu\nu}E_{\mu\nu}(\psi)=0$, 
contains a term $m^2 \psi$, which when combined with another scalar 
equation, $\nabla^{\mu} Q_{\mu}(\psi)= 0$, provides the  
necessary fifth constraint, $\psi =0$. In the present case the trace 
of (8) is 
$$g^{\mu\nu}E_{\mu\nu}(\psi, g)= -\psi_{;\mu}^{\;\; ;\mu}
+\psi^{\mu\nu}_{\;\;\; ;\mu\nu}$$
and no linear combination of this 
equation with the other one, $\nabla^{\mu} Q_{\mu}(\psi, g)= 0$, can 
provide an additional constraint. There is no rigorous proof that the 
fifth constraint does not exist. According to general theory of constrained 
systems of partial differential equations \cite{Pom} some constraints 
may appear only after applying derivatives of a very high order to the 
equations. However it seems unlikely that it may exist.\\

It is well known that in Minkowski space Wentzel Lagrangian, the second 
order action (4) and $G^L_{\mu\nu}(\eta, \psi)$ are gauge-invariant 
under the field transformations 
$\psi_{\mu\nu}\to \psi_{\mu\nu}+\delta\psi_{\mu\nu}=\psi_{\mu\nu}+
\xi_{\mu,\nu}+\xi_{\nu,\mu}$, where $\xi_{\mu}$ is an arbitrary 
vector field. This invariance allows one, as in the case of Maxwell 
field, to introduce five gauge conditions which work then as 
required secondary constraints. This fundamental feature of the 
field is lost in a generic spacetime. Under an infinitesimal transformation 
$\delta\psi_{\mu\nu}=\xi_{\mu;\nu}+\xi_{\nu;\mu}$ the action (4) 
varies by 
$$\delta_{\xi}S_{II}[\psi]= -2\int d^4x\sqrt{-g}\xi^{\mu}Q_{\mu}
\eqno(12)$$ 
plus a surface term which vanishes if $\xi^{\mu}=0$ on the boundary. 
Here $Q_{\mu}=\nabla^{\nu}E_{\mu\nu}(\psi, g)= -\nabla^{\nu}
G^{AD}_{\mu\nu}(\psi, g)$. One sees that a condition for gauge invariance 
of the action and the secondary constraints are directly connected. On the 
other hand the gauge variation of the field equations is different,
$$\delta_{\xi}E_{\mu\nu}= -\delta_{\xi}G^{AD}_{\mu\nu}=
-(R_{\mu\nu ;\alpha}\xi^{\alpha}+ R_{\mu\alpha}\xi^{\alpha}_{\;\; ;\nu}
+ R_{\nu\alpha}\xi^{\alpha}_{\;\; ;\mu}- g_{\mu\nu}R^{\alpha\beta}
\xi_{\alpha ;\beta}-
{1\over 2}g_{\mu\nu}R_{;\alpha}\xi^{\alpha}).\eqno(13)$$
This variation is independent of $\psi_{\mu\nu}$ (the field equations 
are linear) and for arbitrary $\xi^{\mu}$ is determined by Ricci tensor. 
Thus the massless theory is gauge invariant if the two independent 
conditions hold: $Q_{\mu}=0$ and $\delta_{\xi}E_{\mu\nu}=0$. The latter 
condition is satisfied only if the spacetime is empty, $R_{\mu\nu}=0$, 
then also $Q_{\mu}$ vanishes. \\

If $R_{\mu\nu}=0$, i.e.\/ the field $\psi_{\mu\nu}$ does not backreact 
gravitationally and there is no other matter, the field is gauge 
invariant and the invariance may be employed to simplify the field 
equations by imposing a gauge condition. As in flat spacetime one assumes 
the harmonic de Donder gauge 
$$(\psi^{\mu\nu}- \frac{1}{2}g^{\mu\nu}\psi)_{;\nu}=0$$ 
and it reduces eqs. (8) to 
$$-2G^{AD}_{\mu\nu}= \psi_{\mu\nu ;\alpha}^{\;\;\;\;\;\;\; ;\alpha} +
2R_{\mu\alpha\nu\beta}\psi^{\alpha\beta} + \frac{1}{2}g_{\mu\nu}
\psi_{; \alpha}^{\;\; ;\alpha} =0.\eqno(14)$$
The trace of these equations forms a propagation equation for the scalar 
$\psi$, $\Box \psi \equiv \psi_{; \alpha}^{\;\; ;\alpha} =0$. A residual 
gauge freedom still remains since the gauge condition is preserved if 
$\xi^{\mu}$ is a solution a wave equation $\Box \xi^{\mu}=0$ (the spacetime 
is empty). Computing divergence of this equation and using several times 
Bianchi identity and $R_{\mu\nu}=0$ give rise to a scalar equation 
$\Box \xi^{\mu}_{\;\; ;\mu}=0$. The residual gauge transformation alters 
the trace of $\psi_{\mu\nu}$ to $\psi' = \psi +2\xi^{\mu}_{\;\; ;\mu}$. 
For $\psi$ and $\xi^{\mu}_{\;\; ;\mu}$ satisfy the same wave equation one 
can choose such a solution $\xi^{\mu}$ that $\psi +2\xi^{\mu}_{\;\; ;\mu}
=0$ and in this gauge $\psi_{\mu\nu}$ is traceless. 
Finally one has five gauge conditions or secondary constraints, 
$\psi^{\mu\nu}_{\;\;\; ;\nu}=0 =\psi$ ensuring the purely spin-2 nature of the 
field and the field equations take on the form 
$$\Box \psi^{\mu\nu} +
2R_{\mu\alpha\nu\beta}\psi^{\alpha\beta} =0.\eqno(15)$$
 These form a nondegenerate 
system of 10 hyperbolic propagation equations. As a final step 
in constructing a consistent dynamics one proves the 
proposition \cite{MSun}: \\
if the equations of motion (15) hold throughout an empty spacetime 
and the following constraints restrict the initial data at $t=0$: 
$E_{0\mu}=0$ and $\psi=\psi_{;0}=\psi^{\mu\nu}_{\;\;\;\; ;\nu}
=0$, then all the constraints, $\psi^{\mu\nu}_{\;\;\;\; ;\nu}
=0=\psi$ and $E_{0\mu}=0$ are preserved in time. 
Here one must assume an additional initial data constraint 
$\psi_{;0}=0$ at $t=0$ to ensure vanishing of $\psi$ in whole  
spacetime.\\
This gauge invariant theory describes a massless spin--2 field and 
is dynamically consistent. However the field cannot self--gravitate nor 
any other self--gravitating matter may be present. From the viewpoint of 
general relativity such a field is unphysical. It is therefore of crucial 
importance whether the theory in a general spacetime, described by equations 
(4) to (8), is dynamically consistent. \\

One conjectures that the theory (4)--(8) provides a "unified" description 
of a mixture of a purely spin--two field and a scalar one and the lack of the 
fifth constraint means that the scalar cannot be removed from the unifying 
quantity $\psi^{\mu\nu}$. It might be so if the four equations $Q_{\mu}=0$ 
were genuine constraints on $\psi^{\mu\nu}$. 
However eqs. (10) are defective in the following 
sense. It is natural to view them as first order differential 
constraints on the initial data for $\psi_{\mu\nu}$ at $t=0$. 
But then replacing $R_{\mu\nu}$ by the energy--momentum tensor $T_{\mu\nu}$ 
for $\psi^{\mu\nu}$ one sees that they take on the  following form: 
$$Q'_{\mu} \equiv -\nabla_{\beta}(\psi^{\alpha\beta}T_{\mu\alpha})+{1\over 2}
\nabla_{\alpha}(\psi^{\alpha}_{\;\;\mu}T)- \frac{1}{4}\nabla_{\mu}(T\psi) +
\frac{1}{2}T_{\alpha\beta ;\mu}\psi^{\alpha\beta} + \frac{1}{2}
\psi_{;\alpha}T^{\alpha}_{\;\;\mu} =0.\eqno(16)$$
The expression for $T_{\mu\nu}$ is extremely complicated (see below) and, 
what is more important, it
contains $\psi_{\mu\nu;\alpha\beta}$, among them there are second time 
derivatives. Hence (16) actually are four 
nonlinear third order propagation equations. As a consequence 
there are no secondary constraints imposed on $\psi_{\mu\nu}$ which are 
preserved in time and which decouple the unphysical modes 
ensuring the existence of the correct number of degrees of freedom.\\ 

   The opposite possibility is to consider $\psi_{\mu\nu}$ as a 
test field on a fixed background determined by Einstein  
equations $G_{\mu\nu}=t_{\mu\nu}(\phi)$ with $t_{\mu\nu}$ being 
the stress tensor for some matter $\phi$. Then $E_{0\mu}=0$ and 
$Q_{\mu}=0$ are constraints on initial values of $\psi_{\mu\nu}$. 
The necessary condition for having a consistent dynamics for 
$\psi_{\mu\nu}$ is that both primary and secondary constraints 
are preserved in time. This can be shown only for very special 
cases \cite{MSun}. In general there is no consistent dynamical description 
of this mixture of spin-2 and spin--0 fields on a given curved spacetime. 
Furthermore this approach is 
flawed by the assumption that the field does not self--gravitate.\\ 

The third possibility is to regard $Q_{\mu}=0$ as restrictions 
on the spacetime metric. From (10) 
one sees that they contain third time derivatives of the metric 
and thus restrict it in the whole spacetime. It seems  
(though there is no rigorous proof) that there is only one admissible 
solution, $R_{\mu\nu}=0$ and then one comes back to the gauge 
invariant case. 

 The final conclusion is \cite{AD1,AD2}: {\em massive linear 
spin-2 field is consistent only if it is a test field in an 
empty spacetime; then in the limit of vanishing mass it coincides 
with small gravitational fluctuations\/}. The same argument 
applies to the massless field. Inclusion of any non-minimal 
coupling to gravity cannot help \cite{AD1}. A linear spin-2 field 
cannot be a source of gravity and in this sense is unphysical. 

\subsection*{5. Symmetries and the metric stress tensor}
It was mentioned in the introduction that the gauge invariant linear spin--2 
field, which may be interpreted as a linearized gravitational perturbation 
around an $R_{\mu\nu}=0$ solution, has a gauge dependent energy--momentum tensor. 
This is a serious drawback of the theory for the lesson we have learnt from 
general relativity 
is that the adequate description of energy and momentum of any kind of matter, 
except for the gravitational field itself, is in terms of the 
variational (with respect to the
spacetime metric) energy--momentum tensor, hereafter denoted as the metric 
stress tensor. 
In the gauge theories of particle physics the metric stress tensors for the gauge
fields are all gauge invariant. This may arouse a conviction that this is a 
generic feature of
any gauge invariant theory. However this is not the case. In general the metric 
stress tensor does not inherit the gauge--independence property of the 
underlying Lagrangian in Minkowski space. 

An answer why it is so follows from a "folk theorem", rigorously stated and proven by
Deser and McCarthy in \cite{DMC} to the effect that the Poincar\'e generators, 
being spatial
integrals of the metric stress tensor, are gauge invariant and thus unique. The
theorem shows that in quantum field theory, where only global quantities, such 
as total energy and momentum of a quantum
system, are physical (measurable) ones, the inevitable gauge dependence of the
metric stress tensor (for fields carrying spin larger than one) is quite
harmless. It follows from the proof that the gauge dependence of this tensor is
due to the fact that the gauge transformations involve the spacetime metric. It is
also stated in that work that this gauge dependence is unavoidable, there is 
no linear spin--2 field in flat spacetime with a gauge invariant stress 
tensor. 

In a classical gauge invariant field theory any gauge dependence of the
metric stress tensor is truly harmful since this tensor cannot act as the source
in Einstein field equations and this defect makes the theory unphysical. Even if
such a field is viewed as a test one in a fixed spacetime, its theory remains
defective since the local conserved currents (which exist if there are Killing
vectors) do not determine physical flows of energy or momentum through a boundary
of a spatially bounded region. \\
It turns out that the gauge fields are not exceptional in the gauge symmetry 
breaking by the metric stress tensor for fields with spins $s>1$. Actually it 
is a generic effect: for a symmetry transformation of the Lagrangian the
metric stress tensor inherits the symetry property provided that either the field
equations hold or the transformation is metric independent \cite{MaSo}. Thus 
the spacetime metric plays a key role for all symmetries in a field theory
and not only for gauge invariance. Here we follow the generic approach to the 
problem given in \cite{MaSo}.\\

Let $\phi$ be a dynamical
field or a multiplet of fields described by a generally covariant action
functional with a Lagrangian density $\sqrt{-g}L(\phi, g_{\mu\nu})$, residing 
in a curved spacetime with a
(dynamical or background) metric $g_{\mu\nu}$; for simplicity we  assume that 
$L$ does not
depend on second and higher derivatives of 
$\phi$. Let $\phi=\varphi(\phi', \xi, g)$ be any invertible 
transformation of the dynamical variable, which in general involves 
the metric tensor 
and a non--dynamical vector or tensor field $\xi$ and its first 
covariant derivative 
$\nabla\xi$. The transformation is arbitrary with the exception that we exclude 
the tensor
(or spinor) transformations of the field under a mere coordinate transformation.
As a consequence, as opposed to many authors we do
not view the transformation group of the dynamical variables induced by spacetime
diffeomorphisms as a gauge group. The transformation need not be infinitesimal. 
Under the change of the dynamical field one sets 
$$L(\phi, g)=L(\varphi(\phi', \xi, g), g)\equiv L'(\phi', \xi, g).\eqno(17)$$
It is convenient to define the variational stress tensor (signature is $-+++$) 
by the following expression, which is equivalent to the standard definition, 
$$\delta_g(\sqrt{-g}L)\equiv -\frac{1}{2}\sqrt{-g}[T_{\mu\nu}(\phi, g)
\delta g^{\mu\nu}+\hbox{div}],\eqno(18)$$
where div means a full divergence which is usually 
dropped; we will mark its presence from time to time to display an exact 
equality. In evaluating the variation in eq. (18) one assumes that 
$\phi$ is a fundamental field, i.e. is {\em not\/} affected by metric 
variations, $\delta_g\phi=0$. This is the case of the vector potential 
(one--form) $A_{\mu}$ in electrodynamics while $A^{\mu}$ is already 
metric dependent with $\delta_g A^{\mu}=A_{\nu}\delta g^{\mu\nu}$. 
Hence, in evaluating $\delta_g L$ one takes into account only the 
explicit dependence of $L$ on $g_{\mu\nu}$ and $g_{\mu\nu,\alpha}$ 
(or covariantly, on $g_{\mu\nu}$ and $\Gamma^{\alpha}_{\mu\nu}$). \\

In terms of the new field $\phi'$ and of the transformed Lagrangian 
$L'$, the stress tensor
of the theory is re-expressed as follows:
$$\delta_g(\sqrt{-g}L'(\phi', \xi, g))\equiv
-\frac{1}{2}\sqrt{-g}\left[T'_
{\mu\nu}(\phi', \xi, g)
\delta g^{\mu\nu}+\hbox{div}\right].\eqno(19)$$
To evaluate $T'_{\mu\nu}(\phi', \xi, g)$ one assumes that the
appropriate
(covariant or contravariant) components of the field $\xi$ are metric
independent, i.e. $\delta_g\xi=0$, while metric variations of the new
dynamical field $\phi'$ are determined by the inverse transformation
$\phi'=\varphi^{-1}(\phi, \xi, g)$, i.e.
$$\delta_{\varphi}\phi'=\frac{\partial \varphi^{-1}}{\partial g^{\mu\nu}}
\delta g^{\mu\nu}+
\frac{\partial \varphi^{-1}}{\partial g^{\mu\nu}{}_{,\alpha}}
\delta g^{\mu\nu}{}_{,\alpha}.\eqno(20)$$
We denote this variation by $\delta_{\varphi}\phi'$ to emphasize that
$\phi'$
and $g_{\mu\nu}$ are {\em not\/} independent fields: the value
$\phi'(p)$
at any point $p$ depends both on $\phi(p)$ and $g_{\mu\nu}(p)$. Any
scalar or tensor function $f(\phi', \nabla \phi', g)$ depends on the
metric both explicitly (including the connection $\Gamma$) and
implicitly
via $\phi'$, therefore its metric variation is determined by the
{\em substantial ({\rm or} total) variation\/} $\overline{\delta}_g$,
$\overline{\delta}_g f\equiv \delta_g f+\delta_{\varphi}f$. 
Here $\delta_g f$ is the variation taking into
account only the explicit metric dependence of $f$, i.e.
$$\delta_g f\equiv\frac{\partial f}{\partial g^{\mu\nu}}
\delta g^{\mu\nu}+
\frac{\partial f}{\partial \nabla\phi'}
\delta_g\nabla\phi'.\eqno(21)$$
On the other hand the variation $\delta_{\varphi}$ takes
into account the metric dependence of $f$ via
$\phi'=\varphi^{-1}(\phi, \xi, g)$, then
$$\delta_{\varphi}f\equiv \frac{\partial f}{\partial \phi'}
\delta_{\varphi}\phi'+\frac{\partial f}{\partial \nabla\phi'}
\delta_{\varphi}\nabla\phi'\eqno(22)$$
with $\delta_{\varphi}\phi'$ given by (20), 
thus $\delta_{\varphi}$ commutes with the covariant derivative $\nabla$.
For a function $f(\phi, \nabla \phi, g)$ the operators
$\overline{\delta}_g$ and $\delta_g$ coincide, i.e.
$$\overline{\delta}_g f(\phi, \nabla \phi, g)=\delta_g f(\phi, \nabla
\phi, g)
\frac{\partial f}{\partial g^{\mu\nu}}\delta g^{\mu\nu}+
\frac{\partial f}{\partial \nabla\phi}\delta_g\nabla\phi.\eqno(23)$$
Accordingly, $\delta_g$ on the l.h.s. of eq. (19) should be replaced by
$\overline{\delta}_g$. \\

The identity (17), which is valid for {\em all\/} $\phi$,
$g_{\mu\nu}$
and transformations $\varphi$, implies
$\delta_g L(\phi, g)=\overline{\delta}_g L'(\phi',\xi, g)$,
what in turn implies the crucial equality
$$T_{\mu\nu}(\phi,g)=T'_{\mu\nu}(\phi',\xi,g);\eqno(24)$$
in general the two tensors depend differently on their arguments. It 
should be stressed that
in this way one has not constructed the stress tensor for a 
different theory
(also described by the Lagrangian $L'$) in which $\phi'$ would represent a new,
metric--independent field variable. Here, one is dealing with field
redefinitions within the same theory, not with transformations 
relating theories which
are dynamically equivalent but different in their physical 
interpretation.

We are interested in finding out a generic relationship between
$T_{\mu\nu}(\phi,g)$ and $T_{\mu\nu}(\phi',g)$. To this end
we explicitly evaluate
$T'_{\mu\nu}(\phi',\xi,g)$ from the definition (19) and then apply
the
equality (24). 
It is convenient to write the transformed Lagrangian as a sum
$$L'(\phi', \xi, g)\equiv L(\phi', g)+\Delta L'(\phi', \xi, g),
\eqno(25)$$
this is a definition of $\Delta L'(\phi', \xi, g)$. This splitting
allows one to obtain $T_{\mu\nu}(\phi',g)$ upon applying
$\overline{\delta}_g$ to this formula. It is furthermore
convenient to
make the inverse transformation in the term $\Delta L'$, then
$$\Delta L'(\phi', \xi, g)=\Delta L'(\varphi^{-1}(\phi, \xi, g),\xi,g)
\equiv \Delta L(\phi, \xi, g),\eqno(26)$$
and this is a definition of $\Delta L(\phi, \xi, g)$. Then
eq.~(25)
takes the form
$$L'(\phi', \xi, g)= L(\phi', g)+\Delta L(\phi, \xi, g).\eqno(27)$$

The Euler operator of Lagrange equations for $\phi$ arising from $L(\phi,g)$ 
is as usual 
$$E(\phi)\equiv \frac{\delta L}{\delta \phi}=\frac{\partial L}{\partial
\phi}-
\nabla \left(\frac{\partial L}{\partial \nabla \phi}\right);\eqno(28)$$
the tensor $E(\phi)$ has the same rank and
symmetry as the field $\phi$. 
We can now evaluate $T'_{\mu\nu}(\phi',\xi,g)$: employing the definition of 
$\overline{\delta}_g$ and equations (23), (24) and (28), after some 
manipulations and dropping
a full divergence one arrives at the fundamental relationship 
\setcounter{equation}{28}
\begin{eqnarray}
\delta g^{\mu\nu}[T_{\mu\nu}(\phi',g)-T_{\mu\nu}(\phi,g)+g_{\mu\nu}
\Delta L(\phi, \xi, g)]
\nonumber\\
-2E(\phi')\delta_{\varphi}\phi'-2\delta_g\Delta L(\phi, \xi,g)&=&0.
\end{eqnarray}
This is an identity (up to a total divergence) valid for any field,
Lagrangian and any field  transformation. 
We remark that if all full divergence terms were kept in the derivation
of the identity, a divergence term would replace zero on the r.h.s. of
eq. (29). However a total divergence cannot cancel the last three
terms on the l.h.s. of the identity since in general these terms
do not sum up into a divergence. Therefore the difference
$T_{\mu\nu}(\phi',g)-T_{\mu\nu}(\phi, g)$ does not vanish
in general. \\

The transformation $\phi\mapsto\phi'$ is a \emph{symmetry
transformation} of the theory (of the Lagrangian) iff
$L'(\phi',\xi, g)=L(\phi',g)+\hbox{div}$, i.e. if
$\Delta L(\phi, \xi,g)=\hbox{div}$ or is zero. Equivalently, 
symmetry implies that
$L(\phi,g)=L(\phi',g)+\hbox{div}$. 
These equalities should hold identically for
a symmetry independently of whether the field equations are
satisfied or not.
   According to the
proposition ``the metric variation of a divergence is another
divergence", adding a covariant divergence to $L(\phi,g)$
does not affect the variational stress tensor; in a similar way one
shows that the variation with respect to the dynamical field $\phi$
of a full divergence gives rise to another divergence, thus the
Lagrange field equations remain unaffected too.  It is worth stressing 
that we impose no restrictions on the transformation
$\varphi$ and on possible symmetries --- they should only
smoothly depend on the components of a vector or tensor field $\xi$; 
discrete transformations, like reflections, are
excluded. The identity (29) has deeper consequences
usually when the transformation
$\varphi$ depends on the spacetime metric (possibly through covariant
derivatives of the field $\xi$). For any symmetry (29) reduces to
$$\delta g^{\mu\nu}[T_{\mu\nu}(\phi',g)-T_{\mu\nu}(\phi,g)]
-2E(\phi')\delta_{\varphi}\phi'=0\eqno(30)$$
since for $\Delta L=\hbox{div}$ the two terms containing $\Delta L$ can 
be combined into a divergence and it can be discarded. As a trivial example, consider
Maxwell electrodynamics. Here
$\phi=A_{\mu}$, $\phi'= A_{\mu}+\partial_{\mu}f$ with arbitrary $f$;
since the gauge
transformation is metric independent, $\delta_{\varphi}\phi'=0$. Then the
term
$E(\phi')\delta_{\varphi}\phi'=\nabla_\nu
F^{\mu\nu}\delta_{\varphi}\phi'$ vanishes giving
rise to the gauge invariance of $T_{\mu\nu}$ independently of Maxwell
equations. \\
Since the term $E(\phi')\delta_{\varphi}\phi'$ is different from zero for 
fields not being solutions and for metric--dependent symmetry
transformations, one arrives at the  conclusion: \emph{the metric
energy--momentum
tensor for a theory  having a symmetry does not possess this
symmetry}. It is only
{\em for solutions\/}, $E(\phi')=E(\phi)=0$, that the energy--momentum
tensor does possess the same symmetry,
$T_{\mu\nu}(\phi',g)=T_{\mu\nu}(\phi,g)$. 
In physics one is mainly interested in quantities built up of
solutions of equations of motion, but from the mathematical
viewpoint it is worth noticing that the symmetry property is {\em not\/}
carried over from $S$ to $\delta_g S$.\\

Now we can return to the gauge invariant spin--2 field and its gauge 
dependent stress tensor. In gauge theories of particle physics the field 
potentials are
exterior forms since the fields
carry spin one. Then the gauge transformations are independent of the
spacetime metric and the
identity (30) implies gauge invariance of the stress tensor
for arbitrary fields,
not only for solutions. Yet it is characteristic for gauge
theories that for integer spins larger than one a gauge
transformation necessarily
involves covariant derivatives of vector or tensor fields \cite{FWF},
giving rise to gauge dependent stress tensors. \\

We know that Wentzel Lagrangian for spin--2 field is gauge invariant in 
empty spacetimes
\footnote{In our previous work \cite{MaSo} there is an erroneous 
statement about this property in sect. 3 and particularly eq. (42) in 
that work is misleading.}, 
i.e.\/ $L_W(\psi,g)=L_W(\psi',g)+\nabla_{\alpha}A^{\alpha}
(\psi,\xi, g)$ under $\psi_{\mu\nu}\mapsto
\psi'_{\mu\nu}\equiv \psi_{\mu\nu}+\xi_{\mu;\nu}+\xi_{\nu;\mu}$  
for some vector $A^{\alpha}$. In a generic curved spacetime the
"gauge" transformation is no more a symmetry since
$$L_W(\psi,g)=L_W(\psi',g)+{\rm div}+\Delta L_W(\psi,\xi,g)\eqno(31)$$
with 
$$\Delta L_W = -\delta_{\xi}L_W = 2\xi^{\mu}Q_{\mu}.\eqno(32)$$ 
This expression either follows immediately from eq. (12) (the second order 
Lagrangian in (4) and $L_W$ are equivalent up to a divergence in any 
spacetime) or may be directly derived from (7). 
 For $\delta_g \psi_{\mu\nu} = 0 = \delta_g \xi_{\mu}$
one gets for this gauge transformation
$$\delta_\varphi \psi'_{\mu\nu}=-2 \xi_\alpha\delta\Gamma^\alpha_{\mu\nu}.
\eqno(33)$$
In this case the fourth term in the identity (29) reads
$$-2E^{\mu\nu}(\psi')\delta_\varphi \psi'_{\mu\nu}=4
E^{\mu\nu}(\psi')\xi_\alpha\delta\Gamma^\alpha_{\mu\nu}$$
and discarding the full divergence arising from $\delta
g^{\mu\nu}{}_{;\alpha}$ one arrives at
the following explicit form of (29) for the linear
massless spin--2 field and the transformation
$\psi'_{\mu\nu}=\psi_{\mu\nu}+2\xi_{(\mu;\nu)}$,
\setcounter{equation}{33}
\begin{eqnarray}
\delta g^{\mu\nu}\lbrace
T_{\mu\nu}(\psi',g)-T_{\mu\nu}(\psi,g)+g_{\mu\nu}
\Delta L(\psi, \xi, g)\nonumber\\
+2\nabla_\alpha[2\xi_{(\mu}E_{\nu)}^\alpha(\psi',g)-\xi^\alpha
E_{\mu\nu}(\psi',g)]\rbrace
\\
-2\delta_g\Delta L(\psi, \xi,g)&=&0.\nonumber
\end{eqnarray}

 One is interested in evaluating this
identity for $R_{\mu\nu}(g)=0$, what will be symbolically
denoted by $g=r$.
One has $\left.\Delta L_W\right|_{g=r}={\rm div}$ while
$\left.\delta_g\Delta L_W\right|_{g=r}\not=0$. 
In fact, from (10) one can write $\Delta L_W =2\xi_{\mu}
P^{\mu\alpha\beta}\psi_{\alpha\beta}$, where $P^{\mu\alpha\beta}$ 
is made up of Ricci tensor and covariant derivative operators. 
 Then $\left.\delta_g\Delta L_W\right|_{g=r} = 2\xi_{\mu}
\delta_g (P^{\mu\alpha\beta}\psi_{\alpha\beta})$ and 
this variation does not vanish in empty spacetimes and even in 
Minkowski space is different from zero.

Let us denote the
expression in square brackets in (34) by
$F^\alpha_{\mu\nu}(\psi',\xi,g)$. For $R_{\mu\nu}=0$ 
the gauge invariance implies for any $\psi_{\mu\nu}$ 
that $E_{\mu\nu}(\psi',r)=E_{\mu\nu}(\psi,r)$ and then
$\left.\nabla_\alpha
F^\alpha_{\mu\nu}(\psi',\xi,g)\right|_{g=r}=\nabla_\alpha
F^\alpha_{\mu\nu}(\psi,\xi,r)$. Assuming that $\psi_{\mu\nu}$ is a
solution in an empty spacetime, $E_{\mu\nu}(\psi,r)=0$, 
one gets that $\nabla_\alpha F^\alpha_{\mu\nu}(\psi,\xi,r)=0$. 
Thus, the indentity (34) reduces for
solutions and for $R_{\mu\nu}=0$, to
$$\delta g^{\mu\nu}\lbrace
T_{\mu\nu}(\psi',r)-T_{\mu\nu}(\psi,r)\rbrace
-2\left.\delta_g\Delta L_W(\psi, \xi,g)\right|_{g=r}=0.\eqno(35)$$

This relationship (not an identity) shows that the stress tensor is
\emph{not} gauge invariant
even in flat spacetime. In other terms, the symmetry properties of
the Lagrangian in Minkowski (or empty) 
space are insufficient for determining symmetry properties of the
metric stress tensor in this spacetime. In this sense classical field 
theory in flat spacetime is incomplete and a complete and logically 
closed theory should be formulated in a generic spacetime. 

Finally we give for completeness the explicit covariant form in 
Minkowski space of the gauge 
dependent metric stress tensor generated by Wentzel Lagrangian,
\setcounter{equation}{35}
\begin{eqnarray}
T_{\mu\nu}^{W}(\psi,\eta)&=&
 -2\psi_{\alpha\beta;(\mu}\psi_{\nu)}{}^{\alpha;\beta}
+\frac{1}{2}\psi_{\alpha\beta;\mu}\psi^{\alpha\beta}{}_{;\nu}                          
+2\psi_{(\mu}{}^{\alpha;\beta}\psi_{\nu)(\alpha;\beta)}
 \nonumber\\
&&
+\frac{1}{4}g_{\mu\nu}\left(-\psi^{\alpha\beta;\sigma}\psi_{\alpha\beta;
\sigma}
+2\psi^{\alpha\beta;\sigma}\psi_{\sigma\alpha;\beta}\right).
\end{eqnarray}
In deriving it one assumes that the covariant derivatives commute.

\subsection*{6. Gauge symmetry and gravitational energy} 
The fact that the stress tensor $T_{\mu\nu}^{W}$ depends on the gauge
was known long ago
\cite{AD1}. More interesting is the problem whether there exists a
Lagrangian $L_K(\psi,g)$, which is
equivalent to $L_W$  at least in Minkowski space, but which generates a
different, gauge--invariant
stress tensor $T_{\mu\nu}^{K}$ in this spacetime. The no--go theorem
stating that such gauge
invariant stress tensor does not
exist was given in \cite{DMC}. The authors of that work did not
publish a detailed
proof and only referred to the underlying "folk" wisdom. According to
S.~Deser,
for all gauge fields (with metric dependent gauge transformations) in
flat
spacetime the manifest covariance of energy--momentum density objects is
incompatible with their gauge invariance, i.e. these objects are
either covariant
or gauge invariant but not both \cite{D}. In fact, one can always remove
(in a
non--covariant way) the non--physical components of the fields, so
that the result
will have no residual gauge dependence; notice however that this is
not the same as
producing a gauge--independent definition in the usual sense.
An alternative direct proof of the no--go
theorem based on the relationship (35) was then given in \cite{MaSo}.

The physical relevance of the linear massless spin--two field
$\psi_{\mu\nu}$ in flat
spacetime stems from the fact that it is dynamically equivalent to
linearized
General Relativity and it is an unquestionable requirement that all 
viable theories of gravity should dynamically
coincide in the weak--field approximation with the linearized GR, i.e. gravitation
should be
described in this limit by the field. Hence this field is closely 
related to the problem
of gravitational energy density: it is applied in the field theory 
approach to
gravitation, according to which gravity is just a tensor field
existing in Minkowski space,
which may be (though not necessarily) the spacetime of the physical world. In
these theories of gravity the metric energy--momentum tensor again
serves as the most
appropriate local description of energy for the field \cite{BG}. The
best and most
recent version of field theory of gravitation given in \cite{BG}
satisfies this and other requirements imposed on any gravity theory. 
However, while the linearized Lagrange
equations of their theory are gauge invariant (as being equivalent to
those for
$\psi_{\mu\nu}$), their metric stress tensor in this approximation shares the defect
of all the
metric stress tensors for $\psi_{\mu\nu}$, i.e. breaks the gauge
symmetry. From the obvious condition
that the stress tensor should have the same gauge invariance as the field
equations, it follows that also this approach to gravity does not furnish
a physically acceptable notion of gravitational energy
density. \\
Here one touches a subtle problem of what is actually measurable in 
gravitational physics. If one views a field theory approach to gravity  
as a different 
theory of gravity then one may claim that the ``gauge''
transformation actually maps one
solution of field equations to another \emph{physically distinct}
solution. Then the energy
density need not be gauge invariant and measurements of energy may be
used to discriminate
between two physically different solutions related by the ``gauge''
transformation (which
should then be rather called a ``symmetry transformation''). The
gravitational
field of \cite{BG} or the spin--2 field with Wentzel Lagrangian would then
be
measurable quantities rather than gauge potentials. If the
transformation of these fields is
not an internal gauge but corresponds to a change of physical state,
this raises a difficult
problem of finding out a physical interpretation of it. Clearly it is
not a transformation between reference frames.

Here we adopt the opposite view that the field theory approach to
gravity is merely an
auxiliary procedure for constructing notions which are hard to define
in the framework of GR.
It is commonly accepted that in the weak field limit of GR the
spacetime
metric is measurable only in a very restricted sense: if two almost
Cartesian coordinate
systems are related by an infinitesimal translation
$x'^\mu=x^\mu+\xi^\mu$, then no experiment
can tell the difference of their metrics while the curvature tensor
has the same components in
both systems. This implies that all different coordinate systems
connected by this
transformation actually represent the same physical reference frame
and from the physical
viewpoint the transformation is an internal gauge symmetry
\cite{MTWW}. Thus, showing a
mathematical equivalence of the corresponding field equations is
insufficient to
achieve compatibility of a given approach to gravity with the linearized
GR. The weak field gravity should be described by a gauge potential.
In consequence, any
gravitational energy density should be a gauge invariant quantity.\\

We conclude that the above no--go theorem closes one line of research of
gravitational energy
density. This makes the quest of this notion harder than previously.

\subsection*{7. Nonlinear massive spin--two field generated by NLG 
theories} 
As mentioned in Introduction, a radically different approach to the notion 
of consistently gravitationally interacting spin--2 field is provided by 
metric nonlinear gravity theories. Unfortunately in these lectures we have 
no time and space to do justice to this theory, we can only signal the basic 
concepts and results. For an almost comprehensive exposition of the subject 
we refer the reader to our paper \cite{MS2} and references therein. 

Dynamical evolution of a Lorentzian manifold $(M, \psi_{\mu\nu})$ is 
determined in the framework of a generic NLG theory by the 
Lagrangian density  
$$L=\sqrt{-\tilde{g}}f(\tilde{g}_{\mu\nu},  \tilde{R}_
{\alpha\beta\mu\nu}(\tilde{g}_{\mu\nu}))$$ 
where $f$ is any smooth 
scalar function. This evolution and particle content of the theory 
is studied using Legendre transformation method \cite{MFF1, JK, MFF2}.
One need not view $\tilde{g}_{\mu\nu}$ as a
physical spacetime metric, actually whether $\tilde{g}_{\mu\nu}$ or its
"canonically conjugate" momentum is the measurable quantity determining all
spacetime distances in physical world should be determined only after a
careful examination of the physical content of the theory, rather than
prescribed {\it a priori}. Formally $\tilde{g}_{\mu\nu}$ plays
both the role of a metric  tensor on $M$ and is a kind of unifying 
field which will be decomposed in
a multiplet of fields with definite spins; pure gravity is described in
terms of the fields with the metric being a component of the multiplet.
Except for Hilbert--Einstein and Euler--Poincar\'e topological
invariant densities the resulting variational Lagrange equations are of
fourth order.
The Legendre transformation technique allows one to deal
with fully generic Lagrangians; from the physical standpoint, however,
there is no need to investigate a generic $f$. Firstly,
in the bosonic sector of low energy field theory limit of string effective
action one gets in the lowest approximation the Hilbert--Einstein
Lagrangian plus terms quadratic in the curvature tensor. Secondly, to
obtain an explicit form of field equations and to deal with them
effectively one needs to invert the appropriate Legendre transformation
and in a generic case this amounts to solving nonlinear matrix equations.
Hindawi, Ovrut and Waldram \cite{HOW2} have given arguments that a
generic NLG theory has eight degrees of freedom and the same particle
spectrum as in the quadratic Lagrangian below, the only known physical
difference lies in the fact that in the generic case one expects 
multiple nontrival
(i.e.~different from flat spacetime) ground state solutions. This 
result can be also
derived from the observation that after the Legendre transformation 
the kinetic terms
in the resulting (Helmholtz) Lagrangian are universal, and only the 
potential terms
keep the trace of the original nonlinear Lagrangian. If the latter is 
a polynomial
of order higher than two in the curvature tensor, the Legendre map is 
only locally
invertible and this leads to multivalued potentials, generating a ground state
solution in each ``branch"; yet the form of the potential could 
produce additional
dynamical contraints, affecting the number of degrees of freedom, 
only in non--generic
cases. The physically relevant Lagrangians in field theory depend 
quadratically  on
generalized velocities and then conjugate momenta are linear functions of the
velocities. For both conceptual and practical purposes it is then 
sufficient to
envisage a quadratic Lagrangian
$$L=\tilde{R} + a\tilde{R}^2 +b\tilde{R}_{\mu\nu}(\tilde{g})
\tilde{R}^{\mu\nu}(\tilde{g}).\eqno(37)$$
In four dimensions the term
$\tilde{R}_{\alpha\beta\mu\nu}\tilde{R}^{\alpha\beta\mu\nu}$ can be 
eliminated via Gauss--Bonnet theorem). The Lagrangian cannot be purely
quadratic: it is known from the case of restricted NLG theories (Lagrangian
depends solely on the curvature scalar, $L=f(\tilde{R})$) that the
linear term $\tilde{R}$ is essential \cite{MS1} and we will see that
the same holds for Lagrangians explicitly depending on Ricci tensor
$\tilde{R}_{\mu\nu}$. The coefficients $a$ and $b$ have dimension
$\textrm{[length]}^2$; contrary to some claims in the literature there are no
grounds to presume that they are of order $\textrm{(Planck length)}^2$
unless the Lagrangian (37) arises from a more fundamental theory (e.g.
string theory) where $\hbar$ is explicitly present. Otherwise in a pure
gravity theory the only fundamental constants are $c$ and $G$; then $a$
and $b$ need not be new fundamental constants, they are rather related
to masses of the gravitational multiplet fields. We assume that the
NLG theory with the Lagrangian (37) is an independent one, i.e. 
it inherits
no features or relationships from a possible more fundamental theory. \\

Such a theory gives rise to a massive nonlinear spin-2 field 
(and a massive scalar field) in two ways. Firstly, one asumes that 
$\tilde{g}_{\mu\nu}$ is the spacetime metric and then 
one can both 
lower the order of the equations of motion and generate additional 
fields describing gravity in a way analogous to replacing Lagrange 
formalism by canonical one in classical mechanics. One introduces 
"canonical momenta" conjugate to Christoffel connection for 
$\tilde{g}_{\mu\nu}$ using Legendre transformations with respect to the 
irreducible parts of Ricci tensor \cite{MFF2}:
$$\sqrt{-\tilde{g}}\pi^{\mu\nu}= {\partial L \over \partial S_{\mu\nu}} 
\hspace{0.7cm} \hbox{and}\hspace{0.7cm} \sqrt{-\tilde{g}}\phi={\partial 
L \over \partial \tilde{R}}\eqno(38)$$
where $S_{\mu\nu}\equiv \tilde{R}_{\mu\nu}-{1 \over 4}\tilde{R} \tilde{g}_{\mu
\nu}$. The fields $\pi^{\mu\nu}$ and $\phi$ turn out to be massive 
and carry spin two and zero respectively. The original Lagrangian 
$L$ is then replaced by a Helmholtz Lagrangian $L_H$ generating 
second order field equations for the triplet of the fields \cite{MFF2}. 
It is remarkable that for $\tilde{g}_{\mu\nu}$ one gets exactly 
Einstein's field equations $\tilde{G}_{\mu\nu}(\tilde{g})=T_{\mu\nu}(\tilde{g}, 
\pi, \phi)$ with a stress tensor for the nongeometric part of the 
triplet which is however indefinite \cite{MSun}. 
In a weak-field limit one recovers the well known Stelle's results 
for a quadratic $L$ \cite{St1}. 

   The second approach is more sophisticated. One assumes that 
$\tilde{g}_{\mu\nu}$ appearing in $L$ is a kind of a unifying field 
and does {\em not\/} coincide with the physical spacetime metric 
and $\tilde{g}_{\mu\nu}$ is not a geometric quantity. The genuine 
measurable metric field should be recovered from $L$ via a Legendre 
transformation \cite{MFF1,JK} 
$$g^{\mu\nu}\equiv \vert \det({\partial L\over \partial\tilde{R}_
{\alpha\beta}})\vert^{-1/2}{\partial L\over \partial\tilde{R}_{\mu\nu}}
.\eqno(39)$$ 
If this transformation can be inverted one expresses the 
canonical "velocity" $\tilde{R}_{\mu\nu}$ in terms of the "positions 
and momenta", $\tilde{R}_{\mu\nu}(\tilde{g})=r_{\mu\nu}(g^{\alpha\beta}, 
\tilde{g}_{\alpha\beta})$. To view $g^{\mu\nu}$ as a spacetime 
metric one assumes that it  
is nonsingular, i.e. $\det(\partial f/\partial\tilde{R}_{\mu\nu})
\not=0$ and $g_{\mu\nu}$ is its inverse. In other terms 
one maps $(M, \tilde{g}_{\mu\nu})$ onto $(M, g_{\mu\nu})$ and from 
now on one treats $\tilde{g}_{\mu\nu}$ as some matter tensor field on the 
spacetime $(M, g_{\mu\nu})$. From now on all tensor
indices will be raised and lowered with the aid of this metric.
At this point, to make the following equations more readable, we alter
our notation and denote the original tensor field $\tilde{g}_{\mu\nu}$ by
$\psi_{\mu\nu}$ and its inverse $\tilde{g}^{\mu\nu}$ by
$\gamma^{\mu\nu}$. Then 
$$\psi^{\mu\nu}=g^{\mu\alpha}g^{\nu\beta}\psi_{\alpha\beta}
\hspace{0.5cm}\hbox{and}\hspace{0.5cm} \gamma^{\mu\alpha}
\psi_{\alpha\nu}=\delta^{\mu}_{\nu}=\psi^{\mu\alpha}\gamma_
{\alpha\nu}.$$ 

The fields
$g_{\mu\nu}$ and
$\psi_{\mu\nu}$ will be referred to as  Einstein  frame, EF
=$\{g_{\mu\nu},
\psi_{\mu\nu}\}$.  For the generic Lagrangian $\psi_{\mu\nu}$ is
actually a mixture of  fields carrying spin two and zero. Notice that for
$f$ as in (37),
$$g^{\mu\nu} =\left\vert\frac{\psi}{g}\right\vert^{1/2}[(1+
2a\tilde{R})\gamma^{\mu\nu} +2b \tilde{R}^{\mu\nu}],\eqno(40)$$
hence for $\psi_{\mu\nu}$ being Lorentzian and close to Minkowski
metric, $g_{\mu\nu}$ is also Lorentzian and close to flat metric
(and thus invertible). This shows the importance of the linear term
in (37). 
Here one meets the subtle problem which frame is physical, the original one 
consisting solely of the unifying field $\psi_{\mu\nu}$ or EF. It was 
argued in \cite{MS1} that energy
density is very sensitive to field redefinitions and thus is a good
indicator of which variables are physical. For a restricted NLG theory,
$L=f(\tilde{R})$, Einstein frame is the physical one \cite{MS1}.  \\

  As in classical mechanics one replaces the Lagrangian by the 
Hamiltonian, 
$$H(g,\psi)\equiv \sqrt{-g}g^{\mu\nu}r_{\mu\nu}(g,\psi)-
(-\det \psi_{\alpha\beta})^{1\over 2}f(\psi_{\mu\nu}, r_{\mu\nu})\eqno(41)$$ 
and then the latter by a Helmholtz Lagrangian 
$$L_H(g,\psi)\equiv \sqrt{-g}g^{\mu\nu}\tilde{R}_{\mu\nu}(\psi)-
H(g,\psi).\eqno(42)$$ 
The Helmholtz action $S_H=\int d^4xL_H$ generates Hamilton 
equations for $g_{\mu\nu}$ and $\psi_{\mu\nu}$ as variational 
Lagrange equations and these are of {\it second order\/}. 
Introducing a tensor being the difference of the Christoffel 
connections for the two tensors, $Q^\alpha_{\mu\nu}\equiv 
\tilde{\Gamma}^\alpha_{\mu\nu}(\psi)-\Gamma^\alpha_{\mu\nu}(g)$, 
and applying the following identity valid for {\em any\/} 
two nonsingular tensor fields \cite{MFF1}, 
$$K_{\mu\nu} \equiv \tilde{R}_{\mu\nu}(\psi)-R_{\mu\nu}(g)\equiv
\nabla_{\alpha}Q^{\alpha}
_{\mu\nu}-\nabla_{(\mu}Q_{\nu)\alpha}^{\alpha}+Q^{\alpha}_{\mu\nu} 
Q^{\beta}_{\alpha\beta}-Q^{\alpha}_{\mu\beta}Q^{\beta}_{\nu
\alpha},\eqno(43)$$
(all covariant derivatives are with respect to $g_{\mu\nu}$)  
one can finally express $L_H$ in the form (up to a divergence term) 
\setcounter{equation}{43}
\begin{eqnarray}
&
L_H=\sqrt{-g}[R(g)+K(g,\psi)-2V(g,\psi)]
\equiv \sqrt{-g}[R(g)+g^{\mu\nu}(Q^\alpha_{\mu\nu}Q^\beta_{\alpha\beta}-
&
\nonumber \\
&
Q^\alpha_{\mu\beta}Q^\beta_{\nu\alpha})] 
-\sqrt{-g}[g^{\mu\nu}r_{\mu\nu}(g,\psi)-({\det \psi\over g})^
{1\over 2}f(\psi_{\mu\nu}, r_{\mu\nu})].\hspace{0.5cm}
&
\end{eqnarray} 
Here $K\equiv g^{\mu\nu} K_{\mu\nu}$ being the quadratic polynomial 
in $Q^{\alpha}_{\mu\nu}$ (one sees from (43) that $K$ contains 
a full divergence term which may be discarded) is 
a kinetic Lagrangian for $\psi_{\mu\nu}$ and is {\em universal\/} 
(is independent of the form of $f$) while the potential $V$ is 
determined by the original Lagrangian. It is a straightforward 
calculation to show that the theory based on $L_H$ is dynamically 
equivalent to that based on $L=\sqrt{-\psi}f$ \cite{MFF1,JK}. 
It is far from being obvious that it is possible to define the 
genuine metric $g_{\mu\nu}$ in such a way that the gravitational 
part of the Helmholtz Lagrangian is exactly equal to the curvature 
scalar. In this sense {\em Einstein general relativity is a 
universal Hamiltonian image (under a Legendre transformation) of 
any NLG theory\/}. 

   The field equations $\delta S_H/\delta g^{\mu\nu}=0$ are just 
Einstein ones, 
\setcounter{equation}{44}
\begin{eqnarray}
&
G_{\mu\nu}(g)=T_{\mu\nu}(g,\psi)\equiv Q^\alpha_{\alpha(\mu;\nu)}-
Q^\alpha_{\mu\nu;\alpha}-Q^\alpha_{\mu\nu}Q^\beta_{\alpha\beta}
+Q^\alpha_{\mu\beta}Q^\beta_{\alpha\nu}+r_{\mu\nu}+
&
\nonumber \\
&
{1\over 2}g_{\mu\nu}g^{\alpha\beta}(Q^\lambda_{\alpha\beta;
\lambda}-Q^\lambda_{\lambda\alpha;\beta}+Q^\sigma_{\alpha\beta}
Q^\lambda_{\sigma\lambda}-Q^\sigma_{\alpha\lambda}Q^\lambda
_{\sigma\beta}-r_{\alpha\beta}).
&
\end{eqnarray} 
This $T_{\mu\nu}$ is the metric stress tensor following 
from (44); the quickest way of deriving it is to apply the identity 
(43). In general the stress tensor is indefinite and the energy 
density is not determined by initial data since it depends on 
$\psi_{\mu\nu;\alpha\beta}$. The kinetic part of it (made up of 
$Q^{\alpha}_{\mu\nu}$) is universal while the potential part is 
determined by $r_{\mu\nu}(g, \psi)$ and does not depend explicitly 
on $f$. This stress tensor is rather complicated, nevertheless 
it is considerably simpler than that for the linear inconsistent 
field previously discussed . 

   Lagrange equations $\delta S_H/\delta \psi_{\mu\nu}=0$ are 
too complicated to be presented here \cite{MS2}.
These are quasi-linear 
second order equations whose "kinetic" part is universal (due 
to (44)). This universality shows that also from a mathematical 
viewpoint there is no need to study 
a generic NLG theory---to find out the physical content of all 
these theories it is sufficient to investigate the simplest 
case: the original $L$ being a quadratic function of the curvature. 
 The particle content is 
the same for all cases: in a weak-field limit of (44) we find 
that $g_{\mu\nu}$ describes the massless graviton (helicity 2) 
while $\psi_{\mu\nu}$ is a mixture of massive spin-2 and 
spin-0 fields. Our results are in agreement with 
Stelle \cite{St1, St2} who studied the quadratic Lagrangian 
(37).\\
   We choose the Lagrangian $L=\sqrt{-\psi}(\tilde{R}+a\tilde{R}^2-
3a\tilde{R}_{\mu\nu}\bar R^{\mu\nu})$ with $a=\hbox{const}>0$ of 
dimension $(\hbox{length})^2$. The fourth-order field equations 
then imply $\tilde{R}=0$. This corresponds to absence of the scalar 
canonical momentum $\phi$ defined in (38), $\phi=\hbox{const}$. 
The same holds in EF for $\psi_{\mu\nu}$ which is subject to one 
algebraic and four differential constraints and these together 
ensure that the field describes purely spin-2 
particles with five degrees of freedom. 
The field cannot be massless, $m^2_{\psi}={1\over 3a}$. \\

The theory of nonlinear spin--2 field is quite 
promising. Lagrange equations in Einstein frame look rather 
formidable at first sight, nevertheless it was possible to find out 
a nontrivial exact solution \cite{MS2} in the spacetime of a plane--fronted 
gravitational wave with parallel rays (a $pp$ wave). The field has a 
ghost--like nature (found long ago in the linear approximation), 
however recent arguments by Hawking and Hertog \cite{HH} undermine 
the common conviction that 
any viable theory of quantum gravity should be unitary and causal, 
i.e. should exclude negative energy solutions and ghosts. Their 
arguments, based on a scalar field model, are not directly related 
to the spin--two field theory studied here. Nevertheless they are 
in accord with conclusions of \cite{MS2} that appearance of ghostlike 
features in the linear approximation to the spin--two field theory 
does not dismiss it. 

\subsection*{8. Final remarks}
The content of my lectures delivered at the School was to some extent 
different from the written version. It turned out that the problem of 
physical interpretation and viability of restricted nonlinear gravity 
theories (Lagrangian $L=f(R)$ without an explicit dependence on Ricci 
tensor) was still a matter of hot debate and it was why I decided to 
devote to it a part of my talks. As I was unable to present new 
arguments on the subject besides those already existing in the 
literature, I have not written down this part of my lectures. I advice  
the interested reader to consult my joint paper with Guido Magnano 
\cite{MS1} and my conference talk \cite{So}.

\end{document}